Non-Equilibrium Thermodynamics Formalism for Marcus cross-exchange electron transfer reaction rates

Richa Sethi and M.V.Sangaranarayanan *

Department of Chemistry

Indian Institute of Technology, Madras

Chennai-600036 India

E-mail sangara@iitm.ac.in

FAX : +91 44 22570545

ABSTRACT

It is shown that Marcus' cross-exchange electron transfer rate constant expression can be derived from Onsager's linear Flux-Force formalism of Non-Equilibrium thermodynamics. The relationship among the Onsager's phenomenological coefficients for cross-exchange electron transfer processes is deduced and the significance of the methodology is pointed out.

1. Introduction

The estimation of cross-exchange electron transfer rate constants using the constituent self-exchange rate constants occupies a pivotal role in the theory of reaction rates in view of its extensive validity [1,2]. The cross-exchange rate constant $k_{12}$ for a redox reaction is related to the self-exchange rate constants $k_{11}$ and $k_{22}$ as

$$k_{12} = (k_{11} k_{12} K_{12} f_{12})^{1/2} W_{12} \qquad (1)$$

where $K_{12}$ denotes the equilibrium constant of the cross-exchange electron transfer process while $f_{12}$ and $W_{12}$ consist of various work terms involving the reactants and products. If it is assumed that $f_{12} = 1$ and $W_{12} = 1$ as is customary, a simplified eqn

$$k_{12} \approx (k_{11} k_{12} K_{12})^{1/2} \qquad (2)$$

is obtained, enabling the estimation of the cross-exchange rate constants without any adjustable parameters. The validity of eqn(2) has been extensively investigated for diverse types of reactions [3] and is particularly valuable when one of the two self-exchange rate constants is difficult to measure [4]

In general, eqn (2) is considered as a linear free energy relation on account of the linear dependence of the activation free energy upon the standard Gibbs free energy change. The original formalism of Marcus [1] leading to eqn (1) is based upon statistical mechanical considerations in conjunction with classical electrostatics; however, several subsequent attempts have been made to analyze the functional dependence of the activation energy on the intrinsic barrier and reaction coordinate vis a vis progress variable. Notable among them are the investigations of Rehm and Weller [5] Agmon and Levine [6] Thornton [7] and Murdoch [8] It is of interest to note that the exponent to which the equilibrium constant in eqn 2 is raised may need not always be equal to 0.5 [9,10]. A noteworthy feature of the Marcus formalism underlying eqn (1) is that its basic premise holds good not only for electron transfer, but also for methyl transfer[11], hydride transfer[12], proton transfer [13] etc.

In this Letter, we demonstrate that the cross-exchange relation (2) has a non-equilibrium thermodynamics perspective when viewed using Onsager's flux-force formalism[14]. Further, the methodology propounded here, indicates that there exists a deeper theoretical basis underlying eqn (2).

2. Non-equilibrium Thermodynamics formalism for cross-exchange electron transfer reactions

The description of chemical kinetic schemes using non-equilibrium thermodynamics concepts has profound significance in so far as it provides a general framework in an unified manner. For example, the importance of fluctuations from equilibrium states and the concept of coupled and non-coupled bio-chemical reactions are elegantly brought about solely from the magnitude of Onsager's phenomenological coefficients [14,15].

2.1 Chemical kinetics description of cross-exchange reaction

Consider the cross-exchange electron transfer reaction represented as

$$A_1^r + B_2^r \xrightarrow{k_{12}} A_1^p + B_2^p \qquad (3)$$

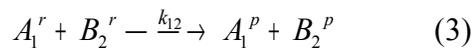

where $A_1$ and $B_2$ represent the two redox couples while r and p denote the reactant and product states. Analogously, the constituent self-exchange reactions are as follows:

$$A_1^r + A_1^p \xrightarrow{k_{11}} A_1^p + A_1^r \qquad (3a)$$

$$B_2^r + B_2^p \xrightarrow{k_{22}} B_2^p + B_2^r \qquad (3b)$$

The equilibrium constant for the cross-exchange reaction is given by

$$K_{eq} = \frac{C_{A_1^p}^{eq} C_{B_2^p}^{eq}}{C_{A_1^r}^{eq} C_{B_2^r}^{eq}} \qquad (4)$$

The velocity of the reaction (3) is

$$v_{12} = k_{12} C_{A_1^r} C_{B_2^r} \qquad (5)$$

where $k_{12}$ is the cross-exchange electron transfer rate constant. In non-equilibrium thermodynamics formalism for chemical kinetics, it is customary to introduce the departure from equilibrium concentrations of the species involved in the reaction. Consequently, eqn(4) becomes

$$v_{12} = k_{12}(C_{A_1^r}^{eq} + \alpha_{A_1^r})(C_{B_2^r}^{eq} + \alpha_{B_2^r})$$

$$= k_{12}[C_{A_1^r}^{eq}(1 + \frac{\alpha_{A_1^r}}{C_{A_1^r}^{eq}}) C_{B_2^r}^{eq}(1 + \frac{\alpha_{B_2^r}}{C_{B_2^r}^{eq}})]$$

$$= k_{12} C_{A_1^r}^{eq} C_{B_2^r}^{eq} (1 + \frac{\alpha_{A_1^r}}{C_{A_1^r}^{eq}} + \frac{\alpha_{B_2^r}}{C_{B_2^r}^{eq}} + \frac{\alpha_{A_1^r} \alpha_{B_2^r}}{C_{A_1^r}^{eq} C_{B_2^r}^{eq}}]$$

Since $v_{12}^{eq} = k_{12} C_{A_1^r}^{eq} C_{B_2^r}^{eq}$, we may write

$$v_{12} - v_{12}^{eq} ; k_{12} C_{A_1^r}^{eq} C_{B_2^r}^{eq} (\frac{\alpha_{A_1^r}}{C_{A_1^r}^{eq}} + \frac{\alpha_{B_2^r}}{C_{B_2^r}^{eq}} + \frac{\alpha_{A_1^r} \alpha_{B_2^r}}{C_{A_1^r}^{eq} C_{B_2^r}^{eq}}) \qquad (5a)$$

For small departures from equilibrium viz. $\frac{\alpha_{A_1^r}}{C_{A_1^r}^{eq}} \ll 1$ and $\frac{\alpha_{B_2^r}}{C_{B_2^r}^{eq}} \ll 1$, the above equation can be approximated as

$$v_{12} ; k_{12} C_{A_1^r}^{eq} C_{B_2^r}^{eq} (\frac{\alpha_{A_1^r}}{C_{A_1^r}^{eq}} + \frac{\alpha_{B_2^r}}{C_{B_2^r}^{eq}}) \qquad (6)$$

This equation is analogous to the velocity expression for a first order reaction [14]

2.2 Onsager's flux-force formalism for cross-exchange electron transfer reactions

In order to obtain insights provided by non-equilibrium thermodynamics, it is customary (reference) to consider the same reaction using the flux-force formalism of Onsager as has been discussed for a first order reaction [15]

2.2.1 Identification of Onsager's phenomenological coefficients

The Affinity of the reaction ( 3 ) is defined as [16]

$$A_{12} = (\mu_{A_1^r} + \mu_{B_2^r}) - (\mu_{A_1^p} + \mu_{B_2^p})$$

where $\mu_{A_1^r}$, $\mu_{B_2^r}$ etc. denote the chemical potentials of the indicated species. Since the liquid phase reactions are considered herein, we may employ the concentrations of the species. Consequently, we write the chemical potential for $A_1^r$ as

$$\mu_{A_1^r} = \mu^0_{A_1^r} + RT \ln C_{A_1^r}$$

if we neglect the activity coefficient corrections. Analogous equations hold good for $\mu_{B_2^r}, \mu_{A_1^p}$ and $\mu_{B_2^p}$ Thus, the Affinity of the reaction becomes

$$A_{12} = [\mu^0_{A_1^r} + RT \ln C_{A_1^r} + \mu^0_{B_2^r} + RT \ln C_{B_2^r}] - [\mu^0_{A_1^p} + RT \ln C_{A_1^p} + \mu^0_{B_2^p} + RT \ln C_{B_2^p}] \quad (7)$$

Introducing the departure from the equilibrium concentrations $\dfrac{\alpha_{A_1^r}}{C_{A_1^r}^{eq}}$ etc. we obtain

$$A_{12} = [\mu^{eq}_{A_1^r} + RT \ln(1 + \dfrac{\alpha_{A_1^r}}{C_{A_1^r}^{eq}}) + \mu^{eq}_{B_2^r} + RT \ln(1 + \dfrac{\alpha_{B_2^r}}{C_{B_2^r}^{eq}})]$$

$$- [\mu^{eq}_{A_1^p} + RT \ln(1 + \dfrac{\alpha_{A_1^p}}{C_{A_1^p}^{eq}}) + \mu^{eq}_{B_2^p} + RT \ln(1 + \dfrac{\alpha_{B_2^p}}{C_{B_2^p}^{eq}})] \quad (8)$$

However, at equilibrium,

$$\mu^{eq}_{A_1^r} + \mu^{eq}_{B_2^r} = \mu^{eq}_{A_1^p} + \mu^{eq}_{B_2^p} \qquad (9)$$

For near-equilibrium conditions, we may expand the logarithmic terms and neglect terms other than linear as is customary in the non-equilibrium thermodynamics description of chemical kinetics [15]. Consequently, eqn (9) becomes

$$A_{12} \cong RT\left(\frac{\alpha_{A_1^r}}{C^{eq}_{A_1^r}} + \frac{\alpha_{B_2^r}}{C^{eq}_{B_2^r}}\right) - RT\left(\frac{\alpha_{A_1^p}}{C^{eq}_{A_1^p}} + \frac{\alpha_{B_2^p}}{C^{eq}_{B_2^p}}\right) \qquad (10)$$

Since the reaction (3) is considered to be an irreversible process (cf. eqn A29 of Marcus [1], the above equation becomes

$$A_{12} \cong RT\left(\frac{\alpha_{A_1^r}}{C^{eq}_{A_1^r}} + \frac{\alpha_{B_2^r}}{C^{eq}_{B_2^r}}\right)$$

Since the velocity is linearly related to the Affinity in the linear flux-force formalism, we may write

$$v_{12} = L_{12} A_{12}$$

However at equilibrium, $A_{12} = 0$; hence $(v_{12} - v_{12}^{eq}) = L_{12} A_{12}$. Hence

$$v_{12} - v_{12}^{eq} = L_{12} RT\left(\frac{\alpha_{A_1^r}}{C^{eq}_{A_1^r}} + \frac{\alpha_{B_2^r}}{C^{eq}_{B_2^r}}\right) \qquad (11)$$

Comparing eqns (6) and (11) we obtain

$$L_{12} = \frac{k_{12} C^{eq}_{A_1^r} C^{eq}_{B_2^r}}{RT} \qquad (12)$$

This equation is reminiscent of the Onsager's coefficient for reversible first order reaction $X \leftrightarrow Y$ with $k_1$ and $k_{-1}$ denoting the forward and reverse rate constants; in this case[17], $L = k_1 C_X^{eq}/RT$. In an analogous manner, Onsager's coefficients for the two self-exchange reactions may be written as

$$L_{22} = \frac{k_{22} C^{eq}_{B_2^r} C^{eq}_{B_2^p}}{RT} \qquad (13) \quad \text{and}$$

$$L_{11} = \frac{k_{11}C_{A_1^r}^{eq}C_{A_1^p}^{eq}}{RT} \qquad (14)$$

2.2 Relation among Onsager's phenomenological coefficients for cross-exchange electron transfer reactions

Since the cross-exchange reaction is composed of the two self–exchange reactions and if the principle of microscopic reversibility is valid in this context,

$$v_{12}^{eq} = v_{11}^{eq} = v_{22}^{eq} \qquad (15)$$

which implies that

$$\frac{k_{12}C_{A_1^r}^{eq}C_{B_2^r}^{eq}}{RT} = \frac{k_{11}C_{A_1^r}^{eq}C_{A_1^p}^{eq}}{RT} = \frac{k_{22}C_{B_2^r}^{eq}C_{B_2^p}^{eq}}{RT} \qquad (16)$$

Thus

$$L_{12} = L_{11} = L_{22} \text{ viz. } L_{12}^2 = L_{11}L_{22} \qquad (17)$$

Substituting the appropriate expressions for L's in terms of rate constants, we obtain

$$k_{12}^2 = k_{11}k_{22}K_{eq} \qquad (18)$$

which is identical with equation (2) – arising from Marcus theory [1].

3. Results and Discussion

The foregoing analysis demonstrates that it is possible to deduce Marcus cross-exchange relation solely from Onsager's Flux-Force formalism when certain approximations are introduced. Interestingly, experimental tests of eqn(2) do indicate satisfactory validity in general [3] and the assumptions made in the present methodology seems reasonable.

It is of interest to enquire whether any new insights have emerged from the approach suggested here. Firstly, in the present version, the departure from equilibrium concentrations of the reactants were explicitly introduced and these were assumed small enabling us to neglect higher order terms in the expansion of the Affinity term in terms of the equilibrium concentrations. A possibility that remains un-clear is whether incorporation of these will yield the complete Marcus expression (1) consisting of the work terms too. Secondly, it is customary [17] to derive and estimate the Onsager's phenomenological coefficient for reversible first order reaction $X \leftrightarrow Y$ wherein the phenomenological coefficient $L = k_1 c_x^{eq}/RT$ where $k_1$ is the first order rate constant in the forward direction, $C_x^{eq}$ being the equilibrium concentration of X. The fact that an analogous exercise is feasible for a cross-exchange reaction (composed of the two self-exchange processes) involving two different reactants may indicate that non-equilibrium thermodynamics formalism is capable of yielding entirely new insights underlying more complicated chemical kinetic schemes. One of the methods of verifying Onsager's Reciprocity Relation (ORR) consists in analyzing a triangular cyclic reaction scheme of coupled chemical reactions. On the other hand, the phenomenological coefficients are related in the present context via eqn(17).

The rate of entropy production of cross-exchange electron transfer processes can be estimated using the above prescription for the Affinity and Onsager's coefficient; this may provide new insights in certain cases. Since there is a correlation between the homogeneous and heterogeneous rate constants, it appears that the present approach can be employed *mutatis mutandis* to heterogeneous chemical kinetics too. In this context, it is worth noting that the Flux-Force formalism of Onsager has been employed to derive Butler-Volmer equation arising in electrode kinetics as shown by Keizer elsewhere[18]. Analogously, the present formalism has led to the analysis of a hierarchy of diffusion-migration equations for electron hopping in redox polymer electrodes [19,20]. This implies that incorporation of the electrochemical potentials in the Affinity expressions would yield a correlation between homogeneous and electrochemical rate constants. It is worth emphasizing that the non-equilibrium thermodynamics formalism applied here provides a pointer to the general validity of Marcus approach for different classes of reactions such as electron transfer, proton transfer, methyl transfer, hydride transfer etc.

4. Conclusions

The linear Flux-Force formalism of Onsager is shown to yield the cross-exchange electron transfer rate constant expression of Marcus.

The phenomenological coefficients are identified and the role played by the (departure from) equilibrium concentrations of the species is indicated.

.